\documentclass{aa}

\usepackage{graphicx}
\usepackage[export]{adjustbox}
\usepackage{xcolor}
\usepackage{float}
\usepackage{bm}
\usepackage{txfonts}

\usepackage[hidelinks=false,pdfusetitle,pdfdisplaydoctitle]{hyperref}
\hypersetup{
    colorlinks=true,   
    linkcolor=blue,    
    filecolor=blue,    
    urlcolor=blue,     
    citecolor=blue     
}

\usepackage{txfonts}
\usepackage{lipsum}
\usepackage{subcaption}         
\usepackage{lscape}             
\usepackage{placeins}

\begin{document}

   \title{Tracing outer planetary systems through white dwarf pollution in wide binaries\thanks{This paper includes data gathered with the 6.5 meter Magellan Telescopes located at Las Campanas Observatory, Chile.}}

   \author{Carolina Lizana-Vidal\inst{1, 2}
          \and Claudia Aguilera-Gómez\inst{1}\and Laura K. Rogers\inst{3} \and Amy Bonsor\inst{4} \and María Ignacia Muñoz-Arriagada\inst{1}\and Patrick Dufour\inst{5} \and Constanza Bravo-Parra\inst{1} \and Ema Salugova\inst{4}}

   \institute{
             Instituto de Astrofísica, Facultad de Física, Pontificia Universidad Católica de Chile, Av. Vicuña Mackenna 4860, 782-0436 Macul, Santiago, Chile\\
             \email{cllizana@uc.cl}
             %\email{craguile@uc.cl}
             \and
             Instituto de Física y Astronomía, Universidad de Valparaíso, Av. Gran Bretaña, 1111, Casilla 5030, Valparaíso, Chile
             \and 
             NOIRLab, 950 N Cherry Ave, Tucson, AZ, 85719, USA
             \and
             Institute of Astronomy, University of Cambridge, Madingley Road, Cambridge CB3 0HA, UK
             \and
             Trottier Institute for Research on Exoplanets and Department of Physics, Universit\'e de Montr\'eal, 1375 Ave. Th\'er\`ese-Lavoie-Roux Montr\'eal, QC H2V 0B3, Canada
             }

   \date{Received March 15, 2026 / Accepted $\langle$ date $\rangle$ }

  \abstract
   {White dwarf (WD) atmospheric metal pollution provides strong evidence for the presence of remnant planetary material and it can be used as an indirect tracer of outer planetary systems.}
   {We analyze whether the presence of a wide stellar companion affects the occurrence of outer planetary systems by comparing the incidence of calcium pollution in hydrogen-atmosphere WDs in wide binaries and in apparently single stars.} 
   {We construct two sample pairs: a large homogeneous SDSS low-resolution spectroscopic set (4844 single WDs; 322 WDs in wide binaries) and a smaller high-resolution archival/targeted sample (83 single WDs; 34 WDs in wide binaries). For each WD, we measure calcium abundances or upper limits and compute detectability-corrected cumulative pollution fractions that account for variations in effective temperature, signal-to-noise ratio, and spectral resolution.}
   {In the SDSS-based samples, we find pollution fractions below $\sim$1\%, defined as the fraction of white dwarfs with detected Ca relative to the total number of white dwarfs, for both single WDs and WDs in wide binaries. After correcting for detectability, the cumulative abundance distributions of the two populations are statistically consistent. The same conclusion is obtained for the higher-resolution samples, despite their different raw detection fractions. These results are subject to limitations including sampling $T_{\text{eff}}$ mismatches, magnitude-dependent selection effects, possible unresolved binaries, He-atmosphere classification difficulties below $\sim13\,000$ K, and coarse model grid precision.}
   {Our results indicate no statistically significant difference in the occurrence rate of remnant outer planetary systems between single stars and stars with wide ($\gtrsim$200 au) companions, within the current uncertainties. These uncertainties are significant because the samples contain only a small number of polluted WDs. The corrected detectability cumulative fraction approach used here provides a framework for comparing samples with different detection sensitivities and can be extended to larger spectroscopic datasets. Such samples will be required to determine whether wide stellar companions affect the survival and delivery of planetary material and to search for more subtle trends, for example, as a function of companion separation.}


   \maketitle
   \nolinenumbers

\section{Introduction}

Planets are common around main-sequence stars. Using a variety of detection techniques, such as transits and radial velocities, it has been established that a substantial fraction of FGK stars host planetary systems \citep[e.g.,][]{WinnFabrycky2015}. At the same time, a significant fraction of stars are not isolated. From $15-50\%$ of solar-type stars are found in binary (or higher-order) multiple systems \citep[e.g.,][]{Raghavan2010, MoeDiStefano2017, Moe2019}.

It is well established that close binary companions can strongly affect planet formation. In tight binaries, typically with separations of a few tens of au or less, tidal interactions truncate the protoplanetary disk \citep{kleyEarlyEvolutionPlanets2010, jang-condellLIKELIHOODPLANETFORMATION2015}, severely inhibiting or in some cases preventing planet formation altogether. Even at wider separations, the presence of a stellar companion can alter the architecture of planetary systems by inducing dynamical perturbations, planet-planet scattering, or migration, ultimately affecting planetary occurrence, eccentricities and inclinations. 

Observationally, planet occurrence shows a strong dependence on the separation of the stellar companion. Exoplanets are less common in close binaries and those with separations below $\sim100$ au. These occurrence rates provide evidence for the suppression of planet formation from close stellar companions \citep{eggenbergerProbingImpactStellar2011,krausImpactStellarMultiplicity2016, hirschUnderstandingImpactsStellar2021, wangInfluenceStellarMultiplicity2014}. In contrast, wide binaries with separations $>1500$ au appear to host planetary systems at rates comparable to those of single stars, suggesting that the influence of the companion becomes weaker at large separations \citep{hirschUnderstandingImpactsStellar2021, wangInfluenceStellarMultiplicity2014, moeImpactBinaryStars2021}. At intermediate separations, however, the impact of binarity remains unclear. Some studies suggest that planet formation may be suppressed by companions at separations as large as $\sim1500$ au, whereas others find no significant effect beyond $\sim200$ au \citep{wangInfluenceStellarMultiplicity2014a,moeImpactBinaryStars2021}. These discrepancies may arise from observational biases inherent to different planet detection techniques, as well as differences between the populations of planets considered in various works (i.e., rocky or giant planets).
Beyond affecting planet occurrence, stellar companions can also influence the detailed architecture of planetary systems, including orbital eccentricities and mutual inclinations \citep[e.g.,][]{AvilaBravo2025} in circumstellar systems, i.e., where the planet orbits only one of the stars. Depending on the orbital parameters of the binary, the characteristics of exoplanetary systems can be significantly different from those of single stars \citep{ThebHagh2015}. 

In the case of P-type orbits, or circumbinary planets, the binary can also have an important dynamical influence on the protoplanetary disk \citep[see ][and references therein]{MarzariThebault2019}, leading to perturbations that can directly modify planet formation \citep{Meschiari2012, Martin2013}. Even after the disk dissipates, the binary can continue to affect the planetary system \citep{ChildsMartin2021}. Observationally, several circumbinary planets have now been confirmed, indicating that planet formation in these environments is possible. Well-studied examples include Kepler-16b \citep{Doyle2011}, Kepler-34b and Kepler-35b \citep{Welsh2012}, among others. However, we focus here on the effects of a stellar companion on circumstellar planets, as we study the influence of wide binaries on planetary systems.
Most observational studies to date have concentrated on identifying stellar companions to known planet-hosting stars, rather than systematically characterizing the presence and architecture of planetary systems in binaries themselves \citep[e.g.,][]{thebaultCompleteCensusPlanethosting2025, messamahMindCompanionDemographics2026}.

In addition to the occurrence rate of exoplanets in binary systems, the incidence of protoplanetary and debris disks around binaries provides a complementary proxy for the presence of planetesimals and planets in these systems.

For protoplanetary disks, the occurrence rate in binaries is strongly dependent on the binary separation. Close binaries, with separations of $\lesssim 100$ au, tend to exhibit significantly lower protoplanetary disk fractions compared to wider systems or single stars \citep{Kraus2012}.
For debris disks, initially \citet{Trilling2007} found that its incidence around main-sequence F–A type binaries is only slightly higher than that observed for single AFGK stars with ages $>600$ Myr, while \citet{Rodriguez2015} could not conclude, with statistical significance, that binary separation affects the presence of debris disks. However, using an updated catalog, \citet{Yelverton2019} showed that no debris disks are found in binaries with separations between $25$ and $135$ au, whereas for wider binaries the occurrence rate rises to $\sim 19^{+5}_{-3}\%$, comparable to that of single stars. This reinforces the idea that binary separation plays a key role in the survival of planetesimals.

While the specific separation thresholds identified across these stages are not directly comparable because of differences in detection techniques, observational biases, and target populations, as well as the fact that binary separations themselves could change over time due to stellar evolution. Nevertheless, studies of protoplanetary disks, exoplanet occurrence, and debris disks consistently show that the presence of a stellar companion has a measurable impact on the survival and architecture of circumstellar material. This raises the question of whether such an impact persists, and in what form, into the final stages of stellar evolution.

As low- and intermediate-mass planet-hosting stars on the main sequence eventually evolve into white dwarfs \citep[WDs, e.g.][]{Lauffer2018new,Althaus2021formation}, these stellar remnants provide a unique opportunity to trace the long-term fate of planetary systems, including those formed in binary environments. 
WDs are characterized by thin atmospheres composed almost exclusively of hydrogen or helium, corresponding to spectral types DA and DB, respectively. Due to their strong surface gravity, heavy elements rapidly sink below the photosphere on short timescales ranging from days in hot WDs to millions of years in cooler ones \citep{KoesterWilken2006, koesterFrequencyPlanetaryDebris2014}. Consequently, the detection of heavy elements in the atmosphere of WDs, known as DAZ, DBZ, or DZ spectral types, provides clear evidence that these objects are externally polluted \citep{zuckermanMetalLinesWhite2003, zuckermanAncientPlanetarySystems2010, koesterFrequencyPlanetaryDebris2014}.

For white dwarfs hotter than $\sim20{,}000$\,K, radiative levitation can counteract gravitational settling and maintain metals in the photosphere \citep[e.g.][]{chayer1995radiative,chayer1995improved}. However, for cooler white dwarfs an external source of metals is required. Early interpretations attributed this pollution to the accretion of material from the interstellar medium \citep{dupuisStudyMetalAbundance1993}, but this scenario has been largely ruled out \citep{farihiRockyPlanetesimalsOrigin2010}. Instead, a wide range of observational and theoretical studies have demonstrated that the dominant source of metals is the accretion of rocky exoplanetary material that survived the evolution of the progenitor star \citep{zuckermanChemicalCompositionExtrasolar2007, farihiRockyPlanetesimalsOrigin2010, dufourDISCOVERYMOSTMETALRICH2010, melisACCRETIONTERRESTRIALLIKEMINOR2011}.

This interpretation is strongly supported by direct observational evidence for remnant planetary systems around white dwarfs. Actively tidally disrupted minor bodies have been detected via transits of disintegrating debris \citep[e.g.][]{vanderburgDisintegratingMinorPlanet2015, rappaportDriftingAsteroidFragments2016, vanderbosch2019white}. In addition, planets and planetary candidates orbiting white dwarfs have been identified through a variety of techniques, including transits, microlensing, direct imaging, and indirect signatures of accretion \citep[e.g.][]{gansickeAccretionGiantPlanet2019, vanderburg2020giant, blackmanJovianAnalogueOrbiting2021, Mullally2024JWST, Limbach2024MEOW}. Further evidence for the accretion of planetary material is provided by the detection of compact dust and gas disks in close proximity to polluted white dwarfs, which are naturally explained as the remnants of tidally disrupted planetesimals that can ultimately supply the observed metal pollution \citep[e.g.][]{Graham1990Infrared,gaensicke2006gaseous}.

WDs offer a complementary approach to traditional planet-detection methods: while transits and radial velocities are most sensitive to close-in planets \citep{winnOccurrenceArchitectureExoplanetary2015}, white dwarfs can probe the existence of outer planetary systems that are otherwise inaccessible to standard detection techniques. Planets at more than a few au are expected to survive the host star's evolution into the WD phase \citep{verasCriticalBinaryStar2018}. Then, even small amounts of planetary material, equivalent to sub-kilometer-sized asteroids or comets, can be dynamically perturbed onto highly eccentric orbits and eventually accreted by the WDs \citep{frewenEccentricPlanetsStellar2014, Farihi2016}. Although small, the presence of this planetary debris can be detected through metal absorption lines in the WD spectrum, indicating recent or ongoing accretion.

Outer planetary systems that have survived stellar evolution to the WD phase appear to be common, with $30-50\%$ of WDs showing some evidence of metal pollution \citep{koesterFrequencyPlanetaryDebris2014, zuckermanAncientPlanetarySystems2010}. A similar fraction of pollution of 30\% has been reported among WDs with very wide ($>1000$au) binary companions, while pollution seems to be rare among WDs in closer binaries ($<1000$au). These conclusions were initially drawn from relatively small samples, such as the 38 WDs with low-mass (M- and K-dwarf) companions studied by \citet{zuckermanOCCURRENCEWIDEORBITPLANETS2014}. \citet{wilsonUnbiasedFrequencyPlanetary2019}, comparing single white dwarfs and wide binaries hosting DA white dwarfs, found similar pollution rates in both populations based on Hubble COS observations, with detection fractions of $45\pm4\%$ for single WDs and $67^{+10}_{-15}\%$ for WDs in wide binaries. More recently, \citet{noorWhiteDwarfPollution2024} estimated a binary fraction of $10.6^{+3.9}_{-3.2}\%$ among polluted DA white dwarfs, based on a sample of 71 DAZ from \citet{wilsonUnbiasedFrequencyPlanetary2019}, and less than $2\%$ among DZ white dwarfs selected from SDSS. While these studies suggest that WD pollution is not primarily driven by the mere presence of a stellar companion, they rely on global pollution fractions and do not explicitly account for differences in detectability across samples. In particular, they do not explore how the inferred pollution rate depends on the abundance level that can be probed in each system.

In this work, we investigate the influence of a wide binary companion on planetary systems by measuring the occurrence rate of polluted white dwarfs, using pollution as a proxy for the presence of outer planetary systems around the white dwarf progenitors. By comparing single and wide-binary white dwarfs using bias-corrected cumulative distributions of Ca detections, used as a tracer of metal pollution given the strength and detectability of the Ca II K line in optical spectra, we account for detectability biases and provide a robust test of whether wide companions affect the occurrence of such systems.

Although the accreted material does not necessarily originate directly from intact exoplanets, their presence in the WD atmospheres is best explained by dynamical perturbations produced by outer planetary companions \citep{DebesSigurdsson2002,Jura2003}. Throughout this work, we interpret the presence of Ca pollution as evidence for the existence of remnant outer planetary systems capable of supplying material to the white dwarf. 
To further investigate the possible role of relatively massive wide companions, we have designed a secondary sample of WDs with FGK solar-type companions, selected from the wide binary catalog of \cite{el-badryMillionBinariesGaia2021}. These solar-type companions may exert a stronger dynamical influence on planetary bodies than low-mass companions, particularly if their orbits evolve to high eccentricities under the action of Galactic tides \citep{kaibPlanetarySystemDisruption2013}. This secondary sample is intended as a complementary exploration of the possible influence of solar-type companions on the presence of Ca pollution, rather than as a direct comparison with the broader wide-binary sample. This raises the question of whether wide binary companions with comparable masses to the actual WD progenitor can destabilize or even destroy planetary systems. 

This paper is structured as follows. Section \ref{sec: data} describes the sample selection of white dwarfs and the observational data. In Section \ref{sec: methods}, we describe how to measure the pollution rates of WDs. Then, in Section \ref{sec: results}, we present our measurements and results on the detectability of pollution, discussing in Section \ref{sec: discussion} the implications and biases present in both single WDs and wide binary systems. Finally, Section \ref{sec: conclusions} summarizes our conclusions.

\section{Sample} \label{sec: data}

This work focuses on a comparison between a control (apparently single) and binary sample. We construct two sets, each divided into two samples of hydrogen-dominated WDs, classified as single and binary systems, and identify which objects show evidence of atmospheric pollution by heavy elements. The first set is based on spectroscopic data from SDSS DR16, which included the largest optical spectroscopic sample of WDs available when this project was initiated. Later data releases of SDSS or DESI could be used to broaden the scope of this analysis in future works. The second set consists of control and wide binary samples observed with different instruments at higher spectral resolution. In this work, we restrict our analysis to DA WDs and exclude DB and other helium-dominated atmospheres. This choice is motivated mainly to minimize observational and modeling biases associated with metal detectability in the atmosphere of polluted WDs, providing a more homogeneous sample for comparing pollution rates across different binary configurations. The two sets are analyzed independently: each wide-binary sample will be compared only with its corresponding control sample, and no direct statistical comparison is performed between Wide Binary samples one and two.

A summary of the samples used in this work and their corresponding sizes is presented in Table~\ref{tab:samples}.

\subsection{SDSS DR16}

\subsubsection{Control sample one}
To compare the pollution fraction in wide binaries with that of single stars, we define a control sample composed of apparently single WDs. Our selection is based on the catalog of WD candidates with available SDSS spectroscopic data by \citet{gentilefusilloCatalogueWhiteDwarfs2021}. In addition to the quality cuts applied in that work, we further impose astrometric constraints and remove known binary systems to minimize contamination from unresolved close or wide companions. This catalog includes estimates of stellar parameters derived from photometric data using Gaia EDR3 \citep{gaiacollaborationGaiaEarlyData2021}.
We first restrict the sample to objects classified as DA, DAZ, and DZA, corresponding to hydrogen-dominated atmospheres with or without detected metals, based on their SDSS optical spectra. Following the criteria adopted by \cite{gentilefusilloCatalogueWhiteDwarfs2021}, all selected sources are required to be high-confidence WD candidates, with probabilities of being a WD $P_{\text{WD}}>0.75$. Given the intrinsic faintness of WDs and their location in the H--R diagram, we further limit the sample to objects with SDSS $g$-band magnitudes brighter than $g<21$.
To avoid contamination from radiative levitation, which can maintain metals in the photosphere of hot WDs \citep{chayer1995radiative,chayer1995improved}, we limit the effective temperature, reported by \cite{gentilefusilloCatalogueWhiteDwarfs2021}, to the range $6000\,\mathrm{K} < T_{\mathrm{eff}} < 20\,000\,\mathrm{K}$. We impose a lower bound of $T_{\mathrm{eff}} > 6000$ K, as below this temperature the Balmer lines become weak, making robust atmospheric parameter and metal abundance determination challenging \citep{Cunningham2025}.

To ensure that objects in the control sample are not affected by unresolved companions, we restrict their Gaia RUWE values to be $<1.4$ and remove sources with confirmed wide binary companions (see Section~\ref{sec:WBS1}). This selection results in a total of 4,879 WDs in Control Sample One, with available SDSS DR16 spectra, before applying the additional cuts described in the following sections.

The \citet{gentilefusilloCatalogueWhiteDwarfs2021} catalog already excludes unresolved close binaries through astrometric and photometric quality cuts, but retains white dwarfs in resolved intermediate and wide binary systems. To ensure a purely single-star control sample, we explicitly remove objects identified as binaries in the \citet{el-badryMillionBinariesGaia2021} catalog.

\subsubsection{Wide Binary sample one} \label{sec:WBS1}

In addition to the selection criteria described in the previous section (spectral type, white dwarf probability, brightness, and effective temperature), we identify wide binaries by cross-matching the catalog of \citet{gentilefusilloCatalogueWhiteDwarfs2021} with the wide binary catalog of \citet{el-badryMillionBinariesGaia2021}, which uses Gaia EDR3 to identify common proper motion pairs. We select binary systems containing at least one WD, whose companion may be either another WD or a non-degenerate star. We also exclude WDs with Gaia RUWE values greater than 1.4 to reduce contamination from unresolved hierarchical triple or higher-order multiple systems. To minimize the possibility of atmospheric contamination due to winds from the stellar companion, we further restrict the sample to binaries with projected separations greater than 200\,au \citep{zuckermanMetalLinesWhite2003, verasCriticalBinaryStar2018}. Applying these criteria yields a final wide binary sample of 324 systems.

\subsubsection{SDSS Spectroscopic Data}
The catalog of \citet{gentilefusilloCatalogueWhiteDwarfs2021} provides information on SDSS optical spectroscopic observations for each source. We retrieve the spectra (with spectral resolution of $R\sim 1800$) and their corresponding redshift measurements from SDSS DR16 \citep{ahumada16thDataRelease2020a}. 
After retrieving the SDSS DR16 spectra, we account for the fact that some WD candidates in the catalog have multiple reported observations. In these cases, we select the spectrum with the highest signal-to-noise ratio, as reported by \citet{gentilefusilloCatalogueWhiteDwarfs2021}. From the resulting 4,879 spectra in Control Sample One, 4,826 WDs are classified as DA, 29 as DAZ, and 24 as DZA. Similarly, within the wide binary sample of 324 WDs, 320 are classified as DA, 2 as DAZ and 2 as a DZA.

The SDSS spectra, originally provided in vacuum wavelengths, are converted to air wavelengths, and shifted to restframe using the redshift value reported in SDSS DR16.

\subsection{Targeted Sample}
In addition to the SDSS sample, which provides a large number of targets but relatively low spectral resolution, we define a second set of control and wide binary samples based on higher-resolution spectroscopic data. 

Following the same procedure as for the SDSS samples, we apply a WD probability cut of $P_{\text{WD}}>0.75$ and restrict the effective temperature range to $6000\,\mathrm{K} < T_{\mathrm{eff}} < 20,000\,\mathrm{K}$. For these higher-resolution samples, we additionally impose a brighter magnitude limit of Gaia $g<17$, ensuring that spectra with enough signal-to-noise (SNR) can be obtained within reasonable exposure times.

\subsubsection{Control sample two}

For the high-resolution control sample of single WDs, we adopt the sample presented by \citet{zuckermanMetalLinesWhite2003}, who obtained high-resolution HIRES spectra for a set of hydrogen-dominated white dwarfs identifying pollution. These WDs are in \cite{gentilefusilloCatalogueWhiteDwarfs2021} catalog and satisfy the $P_{\text{WD}}$ and magnitude criteria. Some objects have $T_{\text{eff}}$ outside the range adopted, and these are retained as part of the literature sample, but excluded from the comparison with Wide Binary Sample two.

\subsubsection{Wide Binary sample two}
The original sample for the high-resolution wide binary observations is based on a cross-match between the filtered WD catalog from \citet{gentilefusilloCatalogueWhiteDwarfs2021} and the wide binary catalog of \citet{el-badryMillionBinariesGaia2021}. As before, we keep only systems with projected separations larger than 200 au. Systems with smaller separations are excluded from all samples, including both the control and wide-binary samples.

We applied a different selection criterion for this sample to probe possible effects of the stellar companion mass. Specifically, we selected only WDs with an FGK-type wide binary companion by restricting the companion absolute magnitude $3 < \mathrm{M_G} < 6$, computed using the Gaia g-band photometry. This selection, based on the wide-binary spectral type, is intended to be a complementary way to explore pollution rates and the possible influence of a solar-type companion, instead of a direct comparison with the pollution rates measured for Wide Binary sample one.

From this selection, we performed an extensive literature and archival search across several observatory archives. Suitable spectra were found in the ESO Archive. For a subset of the remaining WDs, selected from the parent sample according to target observability and observing constraints, we obtained high-resolution spectra from different facilities, as detailed below.
The full details of all observations of DA white dwarfs, including those retrieved from archives, are listed in Table \ref{table:high_res_observations}.

\paragraph{Archive spectra}

Using the Gaia source identifiers of the selected WDs, we searched for available spectra in the ESO Archive\footnote{\url{https://archive.eso.org/wdb/wdb/adp/phase3_spectral/form}}, including data from the ESPRESSO, GIRAFFE, HARPS, UVES, and X-Shooter instruments.

In all cases, we verified that the spectra correspond to the WDs, that the target names matched known WD identifiers, and that the objects were classified as DA, DAZ, or DZA. From the archive search, we identified 2 WDs observed with X-Shooter and 15 with UVES.
The signal-to-noise ratios reported in Table \ref{table:high_res_observations} correspond to the values measured in the continuum near the Ca~II K line.
\paragraph{X-Shooter at VLT UT3}
A total of 17 H-dominated WDs were observed as part of ESO programme 0112.D-0073(A) (PI: Aguilera-Gómez) at the Paranal Observatory during 2023, using Unit Telescope 3 (UT3) of the Very Large Telescope (VLT) and the X-Shooter spectrograph \citep{vernetXshooterNewWide2011}. Observations were carried out using a $1''\times11''$ slit.
X-shooter provides simultaneous wavelength coverage in three arms.
The UBV arm spans 3000\,\AA\, to 5595\,\AA, with a resolving power of $R\sim5400$, encompassing the Ca II K line and therefore used as the basis of our analysis. The VIS arm (5595–10,240 \AA) data are available but not used here, as this spectral range does not contain additional features relevant to this work. The NIR arm data was obtained in stare mode, and thus are not optimal for further consideration.
We use the reduced spectra retrieved directly from the ESO Phase 3 data products.
\paragraph{MIKE at the Magellan Telescopes}

We obtained spectra of 8 H-dominated WDs using the 6.5m Clay Telescope at Las Campanas Observatory in 3 different runs (Program IDs CN2021B-56, CN2022B-13, and CN2023A-26, PI:Aguilera-Gómez), using the Magellan Inamori Kyocera Echelle (MIKE) spectrograph \citep{bernsteinMIKEDoubleEchelle2003}. Observations were performed with a $0.7''\times5.0''$ slit and $2\times2$ binning, providing a spectral resolution of $R\sim42,000$ in the blue arm and $R\sim33,000$ in the red arm. The data were reduced using the CarPy MIKE pipeline \citep{kelsonEvolutionEarlyTypeGalaxies2000, kelsonOptimalTechniquesTwodimensional2003}.

\begin{figure}
   \centering
   \includegraphics[width=\hsize]{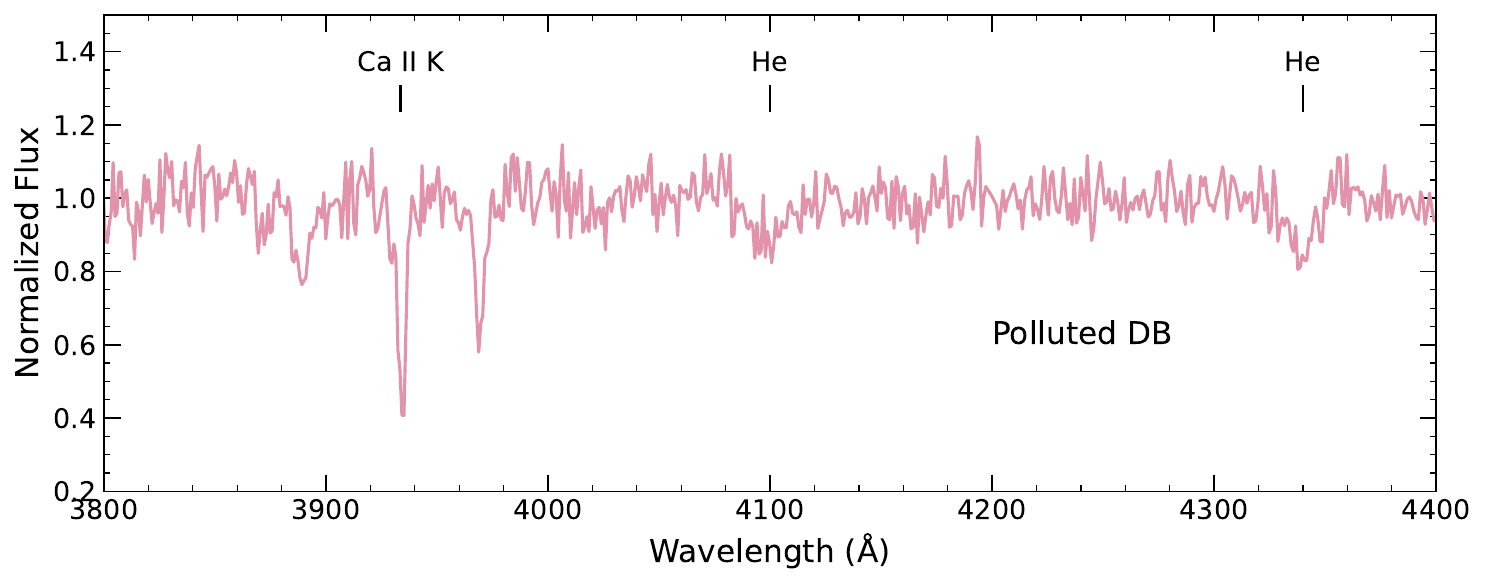}
      \caption{Spectrum of the He-dominated WD Gaia EDR3 836507626075569024, showing the Ca II K line at 3933\AA\ and He lines.}
         \label{fig: DB-example}
   \end{figure}

\begin{figure*}
\sidecaption
    \includegraphics[width=12cm]{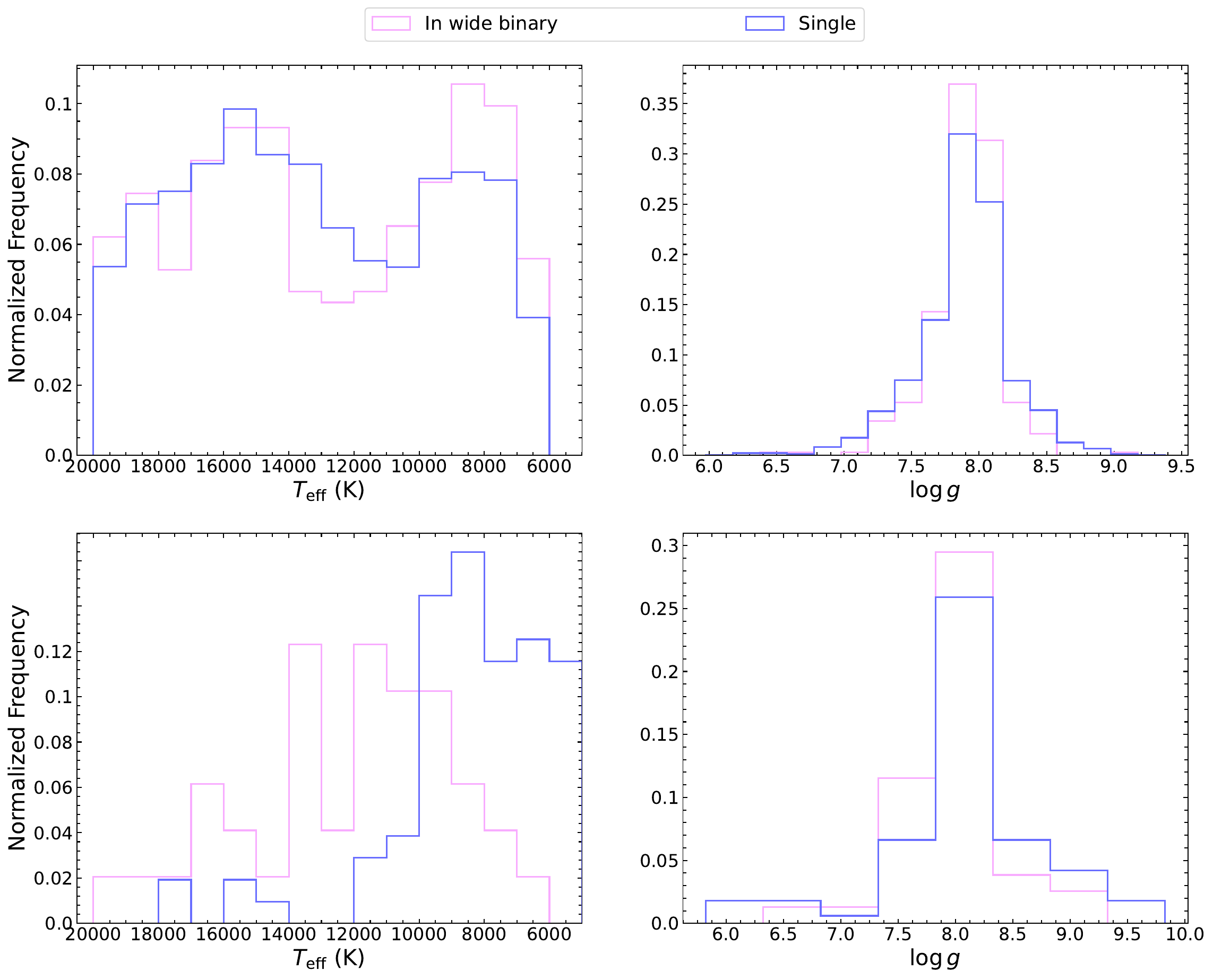}
      \caption{Normalized distributions of $T_{\text{eff}}$ (left) and $\log g$ (right) for the DA WDs in samples analyzed in this work. Top panels: Control Sample 1 (purple) and Wide Binary Sample 1 (pink), both based on SDSS data. Bottom panels: Control Sample 2 from \citet{zuckermanMetalLinesWhite2003} (purple) and Wide Binary Sample 2 (pink).}
    \label{fig: hist_Teff_logg}
\end{figure*}

\subsection{Cleaning the sample from He-dominated WDs}\label{sec: hecleaning}
As our analysis focuses exclusively on DA white dwarfs, we removed from all samples any objects previously identified in the literature as having He-dominated atmospheres, prior to measuring metal abundances. We cross-matched our sources against published catalogs reporting [Ca/He] abundances or helium-rich classifications \citep{dufourSpectralEvolutionCool2007, keplerNewWhiteDwarf2016, keplerNewWhiteDwarf2015, hollandsCoolDZWhite2017, hollandsCoolDZWhite2018, coutuAnalysisHeliumrichWhite2019, vincentClassificationParameterizationLarge2024, vincentDatadrivenSelectionSpectral2023, kilicDetailedModelAtmosphere2026a,kilicDetailedModelAtmosphere2026}.

In addition, we visually inspected each spectrum to confirm the presence of H lines, keeping only sources where these lines were clearly visible. We also examined the spectra for He lines in order to exclude objects showing helium features (for example see Figure \ref{fig: DB-example}). The reported effective temperature was also considered to identify cases in which helium lines might not be visible due to temperature effects, rather than the absence of helium in the atmosphere.

In some cases, we found sources classified as DC white dwarfs in the literature. These objects were also visually inspected, and no spectral lines were identified, confirming their DC nature. They were therefore excluded from the sample.
Moreover, during the spectral inspection and abundance analysis, we identified two objects in SDSS Control Sample 1 whose line profiles and blended Ca features were inconsistent with pure hydrogen atmospheres. Subsequent literature checks confirmed that at least one of these objects has a helium-dominated atmosphere with traces of hydrogen \citep[e.g.,][]{kilicAllskySurveyWhite2025}. As our analysis is restricted to DA WDs, these objects were removed from the final sample.

After this procedure, we obtain the final cleaned samples used throughout this work, which are summarized in Table \ref{tab:samples}. The distributions of effective temperature and surface gravity for both Control Samples and Wide Binary Samples are shown in Figure \ref{fig: hist_Teff_logg}.

\begin{table}[h]
\caption{Summary of the DA WD samples used in this work.}
\label{tab:samples}
\centering
\begin{tabular}{l r}
\hline\hline
Sample & Size \\
\hline
Control Sample 1 (SDSS DR16) & 4\,844 \\
\quad Upper limits (DA) & 4\,821\\
\quad Detections & 23 \\
\hline
Wide Binary Sample 1 (SDSS DR16) & 322 \\
\quad Upper limits (DA) & 320 \\
\quad Detections & 2 \\
\hline
Control Sample 2 \citep{zuckermanMetalLinesWhite2003} & 83 \\
\quad Upper limits (DA) & 62 \\
\quad Detections & 21 \\
\hline
Wide Binary Sample 2 (Targeted Sample) & 34 \\
\quad Upper limits (DA) & 32 \\
\quad Detections & 2 \\
\hline\hline
\end{tabular}
\tablefoot{ For each sample, we report the total number of objects and the number of polluted white dwarfs (detections), as well as objects with no detected metals (upper limits).}
\end{table}

\begin{figure}
   \centering
   \includegraphics[width=0.8\hsize]{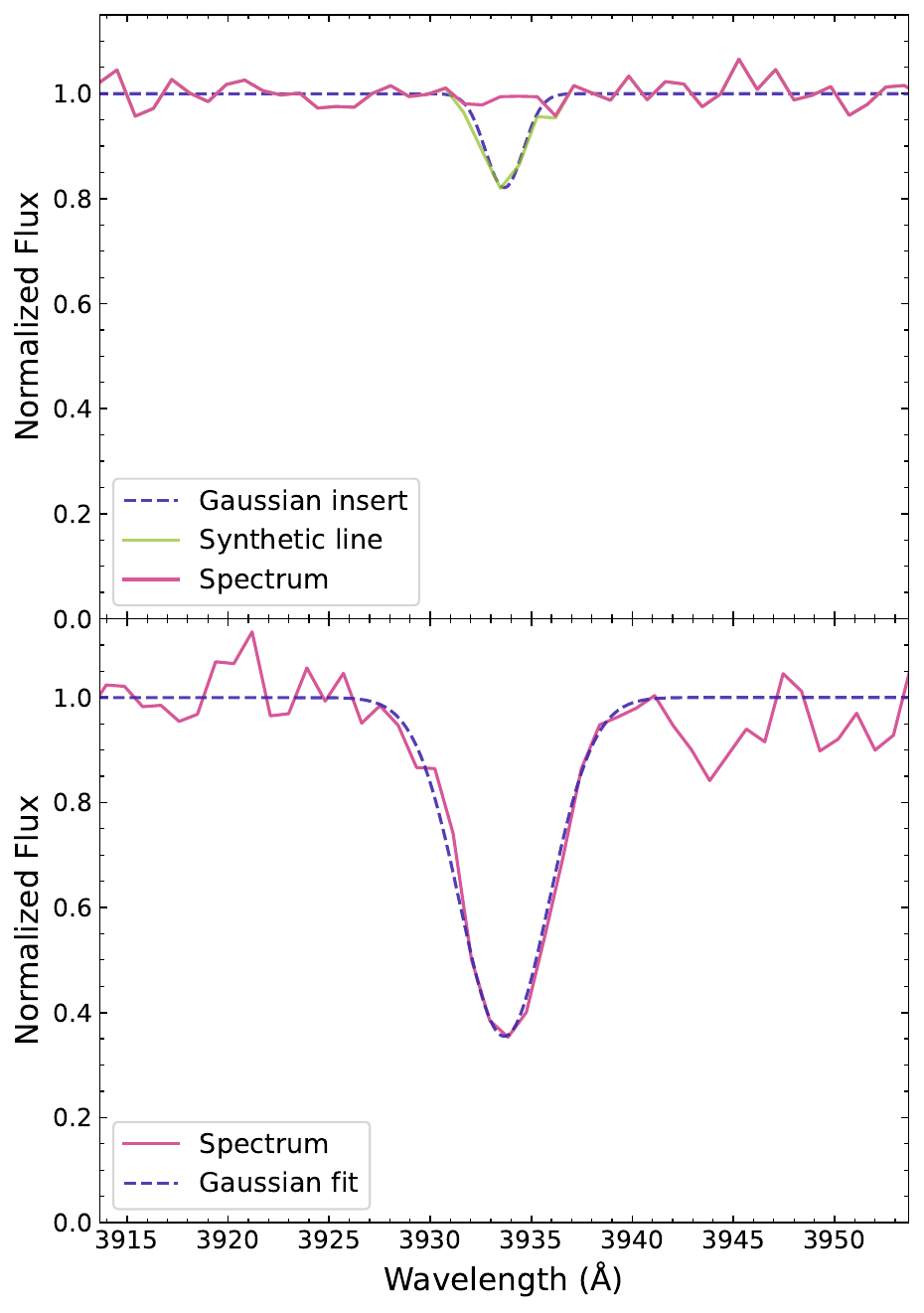}
      \caption{Upper panel: Example of a synthetic Ca II K line absorption line injected into a WD spectrum at a $3\sigma$ significance level relative to the local continuum, at a resolution of $\text{R}\sim1800$. This shows the procedure used to estimate upper limits. Lower panel: Gaussian fit to a detected Ca II K line used to measure its EW shown for the WD Gaia EDR3 2564945432560219008. The spectrum corresponds to the SDSS sample ($\text{R}\sim1800$ and SNR$\sim$25.)}
         \label{fig: lines_method}
   \end{figure}

\section{Method} \label{sec: methods}
The goal of our analysis is to compare the occurrence rate of  pollution in WDs belonging to control and wide binary samples, using metal pollution as a proxy for the occurrence of outer planetary systems.

\subsection{Ca measurements and upper limits} \label{sec: measurements}

Polluted WDs are identified by the presence of metal absorption lines in their spectra. In the optical, the most prominent features are usually the calcium Ca II  H and K lines, although the Ca II H line can be blended with a hydrogen absorption feature. For this reason, we use the Ca II K at $393.37$ nm as our primary tracer of pollution. We searched for Ca II K absorption at the expected wavelength in all spectra in our samples and, for each object, determined either a calcium detection or an upper limit on the equivalent width (EW) of the expected line.

We first derived a $3\sigma$ upper limit on the Ca II K EW for all the spectra, following the method in \citet{Rogers2024SevenI}. We injected a synthetic Gaussian line at the expected wavelength of Ca II K, adopting a width set by the instrumental resolution and noise properties matching the local SNR. The injected strength of the line was progressively reduced, and its detection significance was evaluated through 10,000 Monte Carlo realizations. The $3\sigma$ upper limit was defined as the EW for which 99.7 percent of the trials yielded a $3\sigma$ detection.

We then assessed the Ca II K wavelength region in each spectrum using the same Gaussian line-profile assumptions adopted for the injected lines. When a significant absorption feature was detected at the expected wavelength, we measured its EW by fitting a Gaussian profile to the line. Only features with a significance above $3\sigma$, based on the local signal-to-noise ratio, were classified as detections. Spectra without a significant Ca II K absorption feature were considered as non-detections and assigned the corresponding $3\sigma$ EW upper limit. An example illustrating both Ca II K upper limit and a Gaussian fit to a detected line is shown in Figure \ref{fig: lines_method}.

All upper limits measured in this work are reported in machine-readable tables. The table format is described in Appendix~\ref{app:UL}.

We measure 23 detections in Control sample 1 from SDSS, reported in Table \ref{table:abundances-sdss-singles}, 2 detections in the Wide binary sample 1, detailed in Table \ref{table:detections-binaries-sdss}, and 2 detections in the Wide binary sample 2. These can be found in Table \ref{table:detections-highresolution}. For Control sample 2, \citet{zuckermanMetalLinesWhite2003} reported 21 detections. A summary of the number of detections in each sample is provided in Table \ref{tab:samples}.

The Ca II K EWs were converted into a calcium abundance [Ca/H] using a coarse grid of DA WDs calculated with the same procedure as used in \cite{Dufour2012Detailed}. The grid spectra have effectively infinite resolution and cover the wavelength range of $3100-4700\,\AA$. The effective temperature ranges from $6000\,\text{K}$ to $20\,000\,\text{K}$ in increments of $500\,\text{K}$. $\log g$ spans $7.0$ to $9.0\,\text{cm s}^{-2}$ in steps of $0.5\,\text{cm s}^{-2}$, while  $\log(\text{Ca/H})$ vary from $-4.0$ to $-10.0$, in increments of $0.5$ dex. For each object, we interpolated within this grid using the effective temperatures and $\log g$ reported by \citet{gentilefusilloCatalogueWhiteDwarfs2021}. The uncertainties in the derived $\log(\mathrm{Ca/H})$ abundances were estimated by quadratically propagating the uncertainties in $T_{\mathrm{eff}}$, $\log g$, and the measured EW.

\begin{table*}
\caption{Ca measurements in SDSS Control Sample 1. }

\label{table:abundances-sdss-singles}      
 \centering
\begin{tabular}{ccccccc}
 \hline \hline
  Gaia EDR3 id & $T_{\text{eff}}$ (K) & $\log g\,\,\text{(cm s}^{-2})$ & SNR & EW (m\AA) & $\log{\text{(Ca/H)}}$ & References \\ 
 \hline

97878078428225920 & 6789 & 8.17 & 12 & $2061$ & $-8.08\pm0.36$  & 18\\ 
703588210852298368 & 7170 & 8.27 & 7 & $4903$ &  & 18\\ 
716504796716020352 & 7445 & 7.94 & 45 & $197$ & $-9.55\pm0.10$ & 3, 5, 14, 16\\ 
853645572081536640 & 6507 & 8.10 & 7 & $4369$ & $-7.55\pm0.63$ & 12\\ 
877597264661549568 & 8465 & 8.24 & 20 & $1198$ & $-7.32\pm0.56$ & 18\\ 
1201900590096642688 & 6594 & 7.51 & 5 & $1777$ & $-8.28\pm0.86$ & 18 \\  
1299314816353108992 & 7521 & 8.04 & 29 & $630$ & $-8.52\pm0.10$ &  7, 12, 13, 15, 17 \\ 
1397791368407211264 & 6968 & 8.04 & 19 & $901$ & $-8.64\pm0.11$ & 7, 12\\ 
1411668854417599104 & 8394 & 8.18 & 9 & $5384$ &  & 1, 4\\ 
1437151926176527744 & 10970 & 8.30 & 8 & $1473$ & $-5.44\pm0.34$ & 1, 4, 7\\ 
1460689760003983232 & 8586 & 8.04 & 32 & $361$ & $-8.3\pm0.11$ & 1, 2, 7, 6, 8, 10, 11, 14, 15, 16 \\
1476735444288379392 & 6775 & 7.87 & 11 & $3194$ & $-7.54\pm0.09$ & 9\\ 
1503805214405748736 & 8029 & 8.00 & 12 & $1232$ & $-7.54\pm0.21$ & 18 \\ 
1565012935076040832 & 7523 & 8.08 & 24 & $1322$ & $-7.93\pm0.11$ & 12 \\ 
2555356080553959168 & 7008 & 8.01 & 25 & $507$ & $-9.10\pm0.10$ & 18\\ 
2637627549202579328 & 6498 & 7.99 & 20 & $1043$ & $-8.74\pm0.17$ & 9, 15\\ 
2756297976627824384 & 6790 & 7.74 & 17 & $2738$ & $-7.72\pm0.38$  & 9\\
3145849762325912960 & 6879 & 7.82 & 13 & $1351$ & $-8.33\pm0.46$ & 18 \\
3212596298589400832 & 10676 & 7.57 & 6 & $3005$ & $-4.67\pm0.72$ &  18\\ 
3675799931525903872 & 11219 & 8.00 & 9 & $4116$ & $-4.16\pm0.31$ & 12 \\ 
3907977484067188096 & 7284 & 7.97 & 15 & $966$ & $-8.34\pm0.10$ & TW\\
3954943428189717504 & 6490 & 8.14 & 8 & $7850$ &  & 12, 15\\ 
3999497872730079616 & 13499 & 8.13 & 11 & $4251$ &  & 18\\ 
\hline
\end{tabular}
\tablefoot{Effective temperatures ($T_{\mathrm{eff}}$) and surface gravities ($\log g$) were adopted from \cite{gentilefusilloCatalogueWhiteDwarfs2021}. The SNR was estimated in the vicinity of the Ca II K line. The reported [Ca/H] abundances were measured by fitting with our sparsely sampled model grid and are listed with their uncertainties (Section \ref{sec: methods}). Due to the discrete sampling of the grid, these abundances may differ slightly from values reported in the literature. Blank entries correspond to cases where the Ca abundance could not be reliably measured; further details are provided in the main text. The last column provides the reference(s) where the source has previously been identified as polluted, while white dwarfs newly reported as polluted in this work are marked as TW ("This Work").}
\tablebib{~(1) \cite{mccookCatalogSpectroscopicallyIdentified1999}; (2) \cite{zuckermanMetalLinesWhite2003}; (3) \cite{kilicDebrisDisksWhite2006}; (4) \cite{eisensteinCatalogSpectroscopicallyConfirmed2006}; (5) \cite{farihiSpitzerIRACObservations2008}; (6) \cite{zuckermanAluminumCalciumrichIronpoor2011}; (7) \cite{kleinmanSDSSDR7White2013}; (8) \cite{keplerNewWhiteDwarf2015}; (9) \cite{limogesPhysicalPropertiesCurrent2015}; (10) \cite{guoWhiteDwarfsIdentified2015}; (11) \cite{bedardMeasurementsPhysicalParameters2017}; (12) \cite{gentilefusilloGaiaDataRelease2019}; (13) \cite{kilic100PcWhite2020}; (14) \cite{mccleeryGaiaWhiteDwarfs2020}; (15) \cite{caronSpectrophotometricAnalysisCool2023}; (16) \cite{obrien40PcSample2024}; (17) \cite{kilic100PcWhite2025}; (18) \cite{gentilefusilloCatalogueWhiteDwarfs2021}.}
\end{table*}

\begin{table*}
\caption{Ca measurements for WDs in Wide Binary sample 1 from SDSS.} 
\label{table:detections-binaries-sdss}     
\centering
\begin{tabular}{c c c c c c c c}        
\hline\hline 
Gaia EDR3 id & $T_\mathrm{eff}$ (K) & $\log g\,\,\text{(cm s}^{-2})$ & SNR & EW (m\AA)  & $\log{\text{(Ca/H)}}$ & Separation (au) & References \\
\hline
2564945432560219008 & 7152 & 7.85 & 25 & $3347$ & $-7.26\pm0.19$ &22996 & 1, 4\\
2731772377633104128 & 8071 & 8.00 & 14 & $1241$ & $-7.50\pm0.10$ & 1242 & 1, 2, 3, 4, 5 \\
\hline
\end{tabular}
\tablefoot{The SNR was estimated in vicinity of the Ca II K line. Effective temperatures ($T\mathrm{_{eff}}$) and surface gravities ($\log g$) were adopted from \cite{gentilefusilloCatalogueWhiteDwarfs2021}, and the projected separation between the WD and its companion was taken from \cite{el-badryMillionBinariesGaia2021}. The last column provides the reference(s) where the source has previously been identified as polluted.}
\tablebib{~(1) \cite{mccookCatalogSpectroscopicallyIdentified1999}; (2) \cite{eisensteinCatalogSpectroscopicallyConfirmed2006}, (3) \cite{kilic100PcWhite2020}; (4) \cite{caronSpectrophotometricAnalysisCool2023}; (5) \cite{kilic100PcWhite2025}.}
\end{table*}

\begin{table*}
\caption{Ca measurements for WDs in Wide Binary sample 2.}            
\label{table:detections-highresolution}     
\centering                          
\begin{tabular}{c c c c c c c c}        
\hline\hline                 
Gaia EDR3 id  &$T_\mathrm{eff}$ (K) & $\log g\,\,\text{(cm s}^{-2})$ & SNR & EW (m\AA) & $\log{\text{(Ca/H)}}$ & Separation (au) & References \\    
\hline
2494399362068211328 & 9415 & 8.00 & 33 & $118$ & $-8.88\pm0.10$ & 1143 & TW \\
2842112183412398336 & 10603 & 8.29 & 128 & $364$ & $-6.98\pm0.11$ & 2915 & TW \\
\hline                                   
\end{tabular}
\tablefoot{The SNR was estimated in vicinity of the Ca II K line. Spectral classification were assigned in this work. Effective temperatures ($T\mathrm{_{eff}}$) and surface gravities ($\log g$) were adopted from \cite{gentilefusilloCatalogueWhiteDwarfs2021}, and the projected separation between the WD and its companion was taken from \cite{el-badryMillionBinariesGaia2021}. White dwarfs newly reported as polluted in this work are marked as TW ("This Work").}
\end{table*}

\subsection{Comparison between samples} \label{sec:fraction}
In order to compare the control sample to the wide binary sample, any biases associated with differences in our ability to detect metals between the stars in each sample need to be removed. This is done by computing a detectability-corrected cumulative fraction, defined as the number of stars with Ca detected above a given level ($N_{[Ca/H]>[Ca/H]_p}$), divided by the number of stars for which Ca could be detected to that same level ($N_{[Ca/H]_{upper}>[Ca/H]_p}$). 
Specifically, for a given plotted abundance threshold $[Ca/H]_p$, we define the fraction
\begin{equation}
    f(>Ca/H)=\frac{N_{[Ca/H] > [Ca/H ]_p }}{N_{[Ca/H ]_{upper}<[Ca/H]_p}},\end{equation}
where $\mathrm{[Ca/H]}$ is the measured Ca abundance and $[Ca/H]_{upper}$ is the upper limit on Ca derived for each star (see Section \ref{sec: methods}). The denominator includes both detections and non-detections, provided that the derived upper limit (or measured abundance, in the case of detections) lies below the considered threshold. This method tells us the fraction of stars for which Ca was detected above a given level, taking into account whether or not Ca could be detected to the given level for each star. 
Detections are defined at the $>3\sigma$ level, while upper limits correspond to $3\sigma$ limits; this choice affects only the classification of marginal detections and does not impact the definition of detectability used in the denominator.

\section{Results} \label{sec: results}

\begin{figure*}
   \centering
   \includegraphics[width=0.8\textwidth]{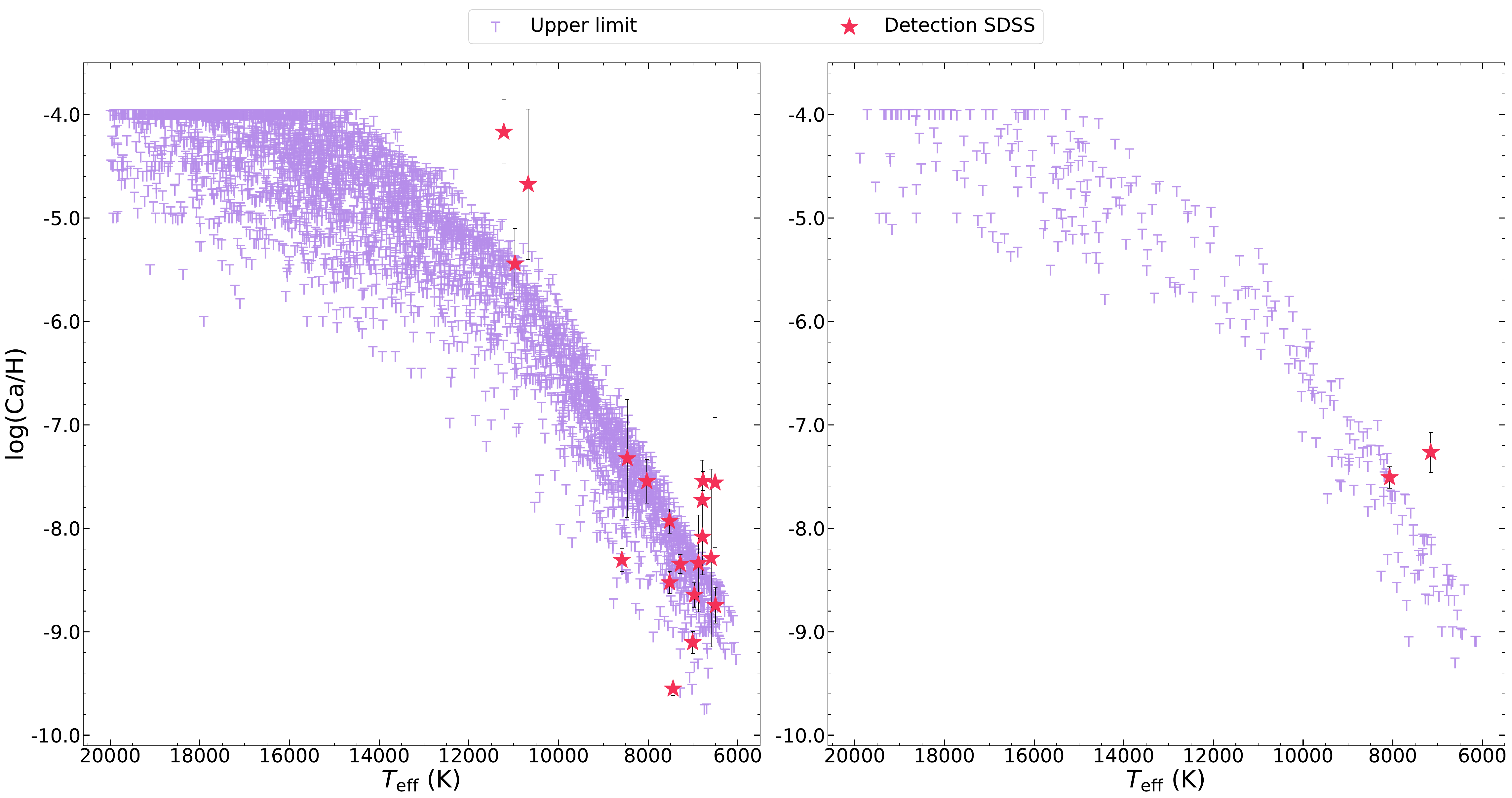}
      \caption{Measured Ca abundances (red star symbol) and upper limits in Control sample 1 (left) and Wide binary sample 1 (right), based on SDSS data.}
         \label{fig: [Ca/H]-SDSS}
   \end{figure*}

Out of the 4,844 DA white dwarfs in Control Sample 1 (SDSS), we report 23 calcium detections. These are found in Table \ref{table:abundances-sdss-singles} except where the star lies beyond the limits of our grid. Taken at face value, this number of detections corresponds to a pollution fraction of $\sim0.47^{+0.12}_{-0.10}\%$. This value is very similar to the fraction measured for Wide Binary Sample 1 (SDSS), where two detections (See Table \ref{table:detections-binaries-sdss}) correspond to a pollution fraction of $\sim0.62^{+0.82}_{-0.41}\%$. These results are based on low-resolution spectra ($R\sim1800$) and are consistent with previously reported metal-pollution fractions of order of a few percent in SDSS white dwarf samples \citep[e.g.,][]{kleinmanSDSSDR7White2013, keplerNewWhiteDwarf2015}. However, a direct comparison of these raw pollution fractions would be misleading, as they do not account for differences in detection sensitivity and sample properties. We therefore report them only as descriptive quantities and to facilitate comparison with previous studies. Our conclusions are instead based on the detectability-corrected cumulative distributions described in Section~\ref{sec:fraction}, which provide a more reliable comparison between the samples.

In contrast, Control Sample 2 from \citet{zuckermanMetalLinesWhite2003}, based on high-resolution spectroscopy, shows a substantially higher raw pollution fraction of $\sim25\%$. Although our Wide Binary Sample 2 is not homogeneous in terms of instrumental setup and spectral resolution, we measure a pollution fraction of $\sim5.88^{+6.7}_{-3.5}\%$, which is significantly higher than that found in the SDSS-based samples. This is based on the 2 detections reported in Table \ref{table:detections-highresolution}. However, a direct comparison between Control Sample 2 and Wide Binary Sample 2 is not straightforward, as the detectability of calcium depends strongly on stellar atmospheric parameters, particularly the effective temperature.

Figure \ref{fig: hist_Teff_logg} (top panels) shows the distributions of effective temperature (left) and surface gravity (right) for Control Sample 1 and Wide Binary Sample 1. Both samples display similar distributions, with comparable $\log g$ values and a bimodal temperature distribution reflecting the presence of both cooler and hotter white dwarfs within our selected temperature range. In contrast, the corresponding distributions for the high-resolution samples (bottom panels of Figure \ref{fig: hist_Teff_logg}) differ significantly. While the $\log g$ distributions of Control Sample 2 and Wide Binary Sample 2 remain similar, the effective temperature distributions are markedly different. In particular, the HIRES sample of \citet{zuckermanMetalLinesWhite2003} is strongly biased toward cooler white dwarfs, for which calcium absorption lines are intrinsically easier to detect. 

\begin{figure}[t]
   \centering
   \includegraphics[width=0.85\hsize]{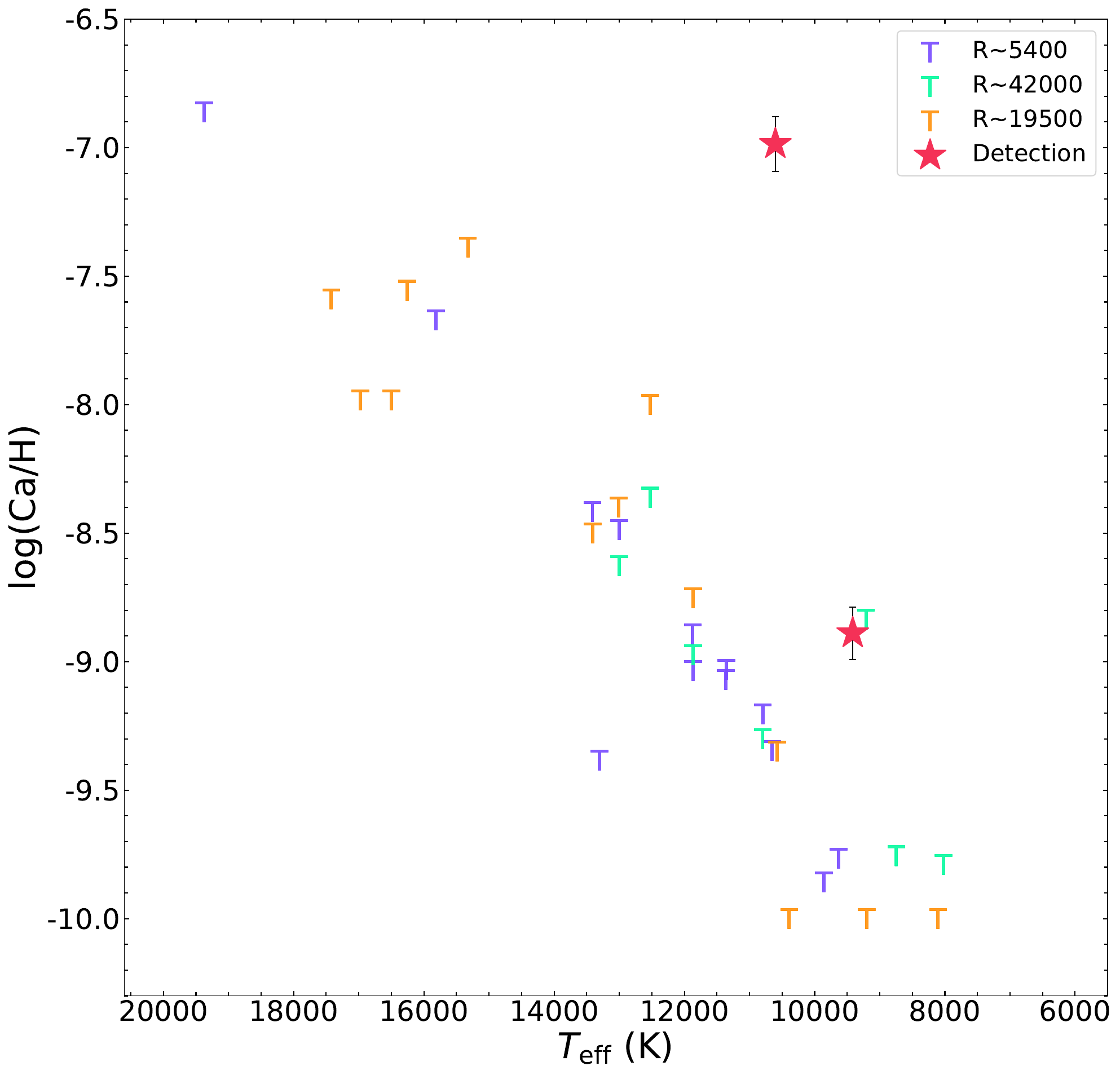}
      \caption{Measured Ca abundances (red star symbol) and upper limits in wide binary sample 2. Upper limits are colored according to the resolution of the data used: X-Shooter R$\sim5400$ in purple, UVES R$\sim19500$ in orange, and MIKE R$\sim42000$ in cyan.}
    \label{fig: [Ca/H]-binaries-highres}
\end{figure}

This temperature dependence is also depicted in Figure \ref{fig: [Ca/H]-SDSS}, which shows that the upper limits on calcium abundance decrease toward lower effective temperatures, implying increased sensitivity to lower calcium abundances in cooler white dwarfs. This trend is observed independently of spectral resolution. Figure~\ref{fig: [Ca/H]-binaries-highres} presents the upper limits and detections for Wide Binary Sample 2, with their resolutions indicated as different colors. As expected, the same temperature dependence is present.

In addition to temperature effects, spectral resolution plays a significant role in calcium detectability. A comparison between Figures \ref{fig: [Ca/H]-SDSS} and \ref{fig: [Ca/H]-binaries-highres} shows that higher-resolution spectra yield systematically lower upper limits on calcium abundance, as absorption features are more easily disentangled from the noise. In consequence, high-resolution data provide stronger constraints on the presence of low-level pollution.

In all cases, most calcium detections lie above the corresponding upper-limit curves, consistent with the expected detectability for stars with similar signal-to-noise ratios.

Given these differences in selection effects, temperature distributions, spectral resolution, and signal-to-noise ratio across the samples, a direct comparison of raw pollution fractions can be misleading. Instead, a more appropriate approach is to compare detectability-corrected pollution fractions, which account for the varying sensitivity to calcium across the parameter space. The procedure used to compute these fractions is described in Section \ref{sec:fraction}.

\subsection{Detectability-corrected comparison}

\begin{figure*}
   \centering
   \includegraphics[width=0.84\textwidth]{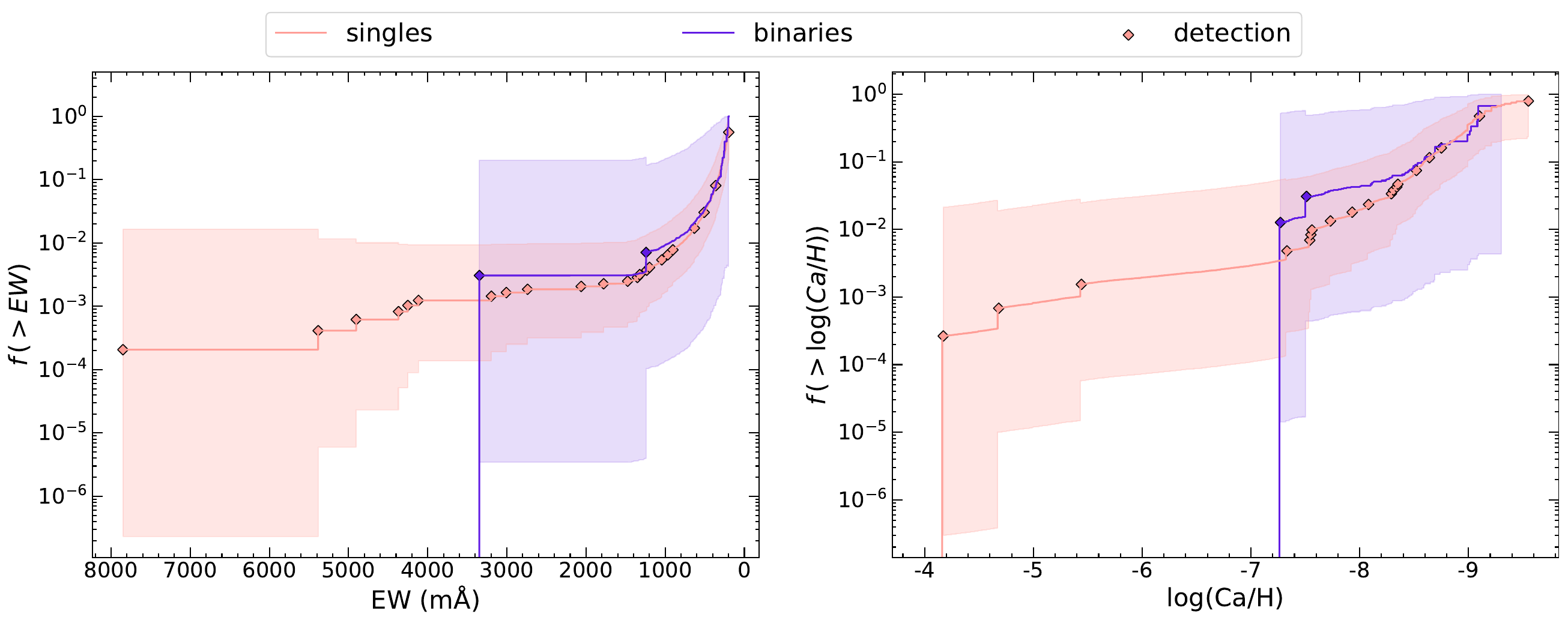}
      \caption{Cumulative fraction of white dwarfs in the SDSS sample with calcium detected above a given threshold, corrected for detection biases using individual upper limits. The Control sample 1 is shown in orange, while the Wide Binary sample 1 is shown in purple. The left panel presents the cumulative fraction as a function of Ca II K equivalent width, and the right panel shows the same quantity expressed in terms of calcium abundance. Shaded regions around each curve indicate the $1\sigma$ confidence intervals.}
         \label{fig: fraction plot SDSS}
\end{figure*}

\begin{figure*}
   \centering
   \includegraphics[width=0.84\textwidth]{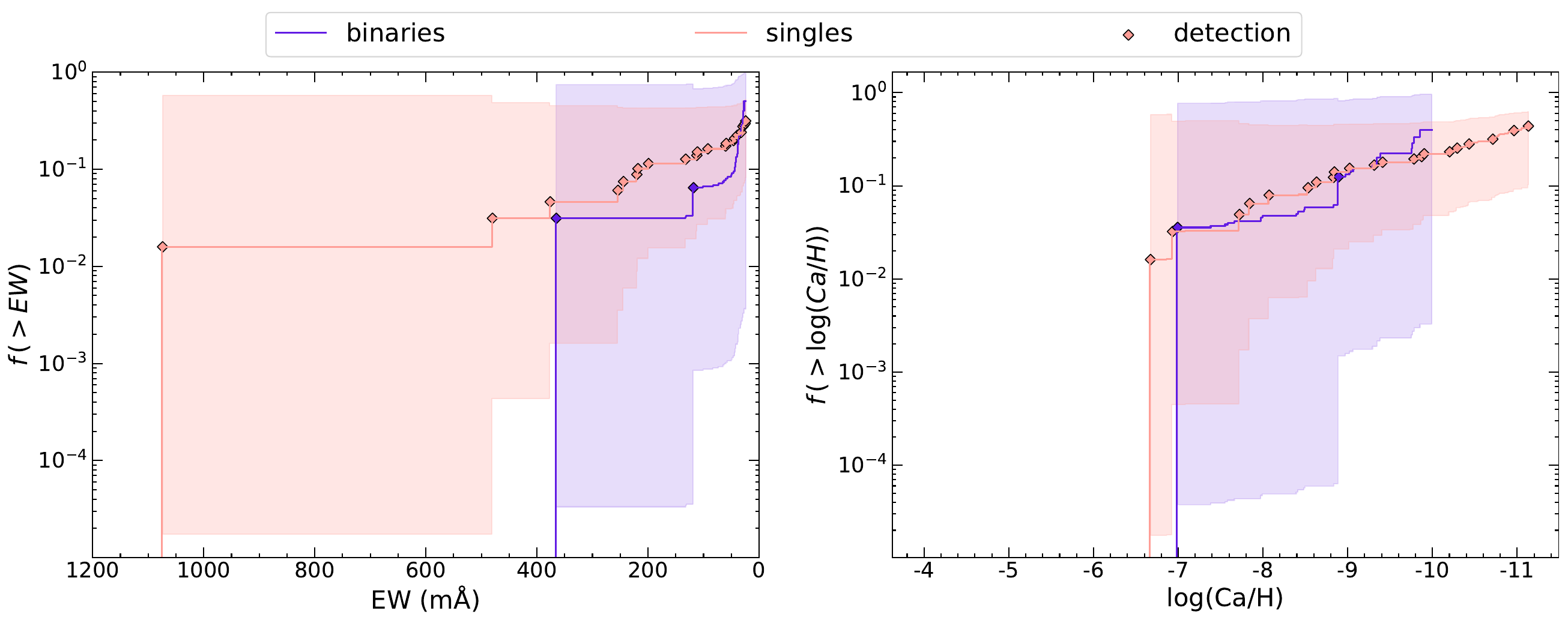}
      \caption{Same as Figure \ref{fig: fraction plot SDSS}, but for Control sample 2 and Wide Binary sample 2.}
         \label{fig: fraction plot targeted sample}
\end{figure*}

The detectability-corrected cumulative fractions of Ca-polluted WDs for each sample are shown in Figure \ref{fig: fraction plot SDSS} for the SDSS-based samples, and in Figure \ref{fig: fraction plot targeted sample} for Control Sample 2 and Wide Binary Sample 2. In all cases, the colored shaded regions represent the $1\sigma$ confidence intervals, estimated using binomial statistics for small-number samples following \cite{gehrelsConfidenceLimitsSmall1986}.

Figure \ref{fig: fraction plot SDSS} shows the cumulative fractions as a function of EW (left panel) and Ca abundance (right panel), comparing single WDs (orange) with white dwarfs in wide binary systems (purple). Within the uncertainties, no significant difference is observed between the two populations over the full range of abundances and EWs probed.
A $\chi^2$ comparison of the cumulative abundance distributions of single and wide-binary WDs in the SDSS sample yields $\chi^2=0.82$, estimated over the common detection range, indicating that both distributions are statistically consistent within the uncertainties. Given the small number of detections in the wide binary sample, the resulting error bars are large, limiting the strength of any potential constraints that can be placed on possible differences between the two distributions. Similar conclusions are obtained for Control Sample 2 and Wide Binary Sample 2, shown in Figure \ref{fig: fraction plot targeted sample}. Over the abundance range where both samples have detections, a $\chi^2$ test yields $\chi^2=1.93$. While this value may suggest a mild discrepancy, it is not statistically significant once the uncertainties on the cumulative fractions are taken into account. We note that some WDs in Control Sample 2 lie outside the effective temperature range adopted for the other samples. However, their reported Ca abundances fall outside the abundance range shared with Wide Binary Sample 2 and were therefore not included in the $\chi^2$ comparison. Consequently, these objects do not affect the statistical comparison presented here.

Overall, after correcting for detectability biases, we find no statistically significant difference in the occurrence rate of Ca pollution between single WDs and those in wide binary systems. While this does not rule out the existence of an intrinsic difference between the two populations, any effect on the occurrence rate of outer planetary systems, as traced by WD pollution, remains undetected with the current samples and their associated statistical uncertainties. Nonetheless, the detectability-corrected cumulative-fraction framework introduced in this work provides an effective methodology for comparing samples with different detection biases and can be applied to larger and more homogeneous datasets to place tighter constraints on this comparison.

\subsection{Newly identified polluted DA WDs}
Some WDs in our samples have not previously been reported as polluted. We mark the only new single SDSS polluted WD in Table \ref{table:abundances-sdss-singles} with a ``TW'' (i.e., This Work)  in the last column, while all stars in Table \ref{table:detections-highresolution} correspond to new detections.  These newly identified polluted WDs provide promising targets for future spectroscopic follow-up and detailed abundance analysis.

In all three newly identified polluted WDs, Ca is the only metal detected in the available spectra. We searched for additional photospheric metal lines but found no significant absorption features. Higher resolution and higher signal-to-noise observations may reveal weaker absorption lines from other elements, enabling a more complete characterization of the bulk composition of the accreted planetary material.

The measured [Ca/H] abundances are consistent with those typically observed in polluted DA WDs at similar effective temperatures, based on the values compiled in the Montreal White Dwarf Database \citep[MWDD,][]{dufourMontrealWhiteDwarf2017}. Therefore, these objects do not appear to have unusually high or low Ca abundances compared with the broader polluted WD population.

\section{Discussion} \label{sec: discussion}

This paper finds suggestive evidence that there is no statistically significant difference in the occurrence of outer planetary systems, witnessed by the presence of Ca pollution in the atmospheres of white dwarfs, between single stars and those with wide binary companions. Although this conclusion is severely limited by the small sample size, it complements conclusions made for main-sequence exoplanets and debris-disk studies indicating that wide stellar companions might not strongly affect planet occurrence.

Although the companions that we consider in this work are all separated by more than $200$ au, such companions retain the potential to influence any planetary system. While direct interactions with the system, including truncation of disks and scattering of planets are prevalent in smaller separation binaries \citep[e.g.,][]{DesideraBarbieri2007,Yelverton2019}, a wider companion has less potential for a direct dynamical influence on an inner planetary system, potentially hundreds of au away. The orbits of wide binary companions, however, can change with time due to the influence of galactic tides. Dynamical scattering during closer incursions of binary companions has been suggested to explain the higher eccentricities of exoplanets in systems with a wide binary companion \citep{kaibPlanetarySystemDisruption2013} and as a way to scatter material close to white dwarfs at late times \citep{bonsorWideBinaryTrigger2015}. For white dwarf planetary systems, the existence of a wide binary companion has the potential to reduce the amount of material that survives in a planetary system to late times (due to dynamical perturbations earlier in the system's evolution) or to enhance scattering at late times. In the latter case, there is a small increased chance of pollution \citep[see][for full details]{bonsorWideBinaryTrigger2015}.

During post-main-sequence evolution, inner planets are generally engulfed during the red giant branch, while outer planets survive and undergo orbital expansion due to stellar mass loss. This mass loss can itself trigger dynamical instabilities among surviving planets \citep{verasPostmainsequencePlanetarySystem2016}, providing a pathway for the inward scattering of minor bodies.

In single-star systems, simulations show that outer planets can efficiently excite instabilities in planetesimal belts and deliver material onto the white dwarf \citep{mustillUnstableLowmassPlanetary2018}. In wide binaries, similar processes may operate, with the companion providing an additional, but not necessarily dominant, source of perturbations \citep{bonsorWideBinaryTrigger2015}. Our results suggest that, for separations $\gtrsim$200 au, this additional perturbation does not produce a measurable change in the overall incidence of pollution.

The coexistence of these competing effects, that is, early depletion versus late-time enhancement of scattering, may naturally lead to a net pollution rate that is similar to that of single stars, consistent with our observational results.

\subsection{Companion separation}
The projected separations in both the SDSS binary sample and Wide Binary Sample 2 span approximately $250$ to $100,000$ au. Given the small number of polluted systems, we cannot draw statistically significant conclusions regarding whether the frequency of pollution depends on companion separation.

Nevertheless, we note a qualitative scarcity of polluted white dwarfs with companions at separations below $\sim1000$ au, similar to trends reported by \citet{zuckermanOCCURRENCEWIDEORBITPLANETS2014} and \citet{williamsPEWDDDatabaseWhite2024}, who found reduced pollution fractions in binaries with separations below $\sim500$ au. If real, this may indicate that moderately wide companions can suppress pollution of DA WDs. Larger samples will be required to test this possibility.
We caution, however, that this apparent scarcity may be influenced by sample selection effects. The \citet{gentilefusilloCatalogueWhiteDwarfs2021} catalog excludes binaries through photometric and astrometric quality cuts, and thus systems with moderately wide separations and luminous companions may be affected by the selections. Thus, it remains unclear whether this apparent deficit reflects a genuine physical effect or is partially driven by selection biases.

\subsection{Limitations}
Our results are subject to some limitations, mainly related to sample selection, properties, and data availability.

\subsubsection{Magnitude limits}
All of our samples are limited by apparent magnitude in the $g$ band, which introduces a bias toward intrinsically brighter WDs. While there is no a priori reason to anticipate that the prevalence of planetary systems depends directly on WD brightness, magnitude limits may indirectly bias the effective temperature distribution of the samples. Any systematic differences in temperature between samples could, in turn, affect the detectability of Calcium. 

\subsubsection{Binarity classification}
The subdivision between single stars and wide binaries is another potential source of uncertainty. Some apparently single WDs may host unresolved or unidentified companions. We rely on the initial selection of \citet{gentilefusilloCatalogueWhiteDwarfs2021}, on RUWE cuts to remove close binaries, and on the binary catalog of \citet{el-badryMillionBinariesGaia2021} to identify common proper-motion pairs, which should remove wide and intermediate separation binaries. Although this catalog has been carefully constructed, chance alignments or missed companions may still occur. Consequently, a small fraction of true binaries could contaminate the control samples, and some single WDs could contaminate the wide binary samples. Residual binarity contamination in the control sample is therefore expected to arise primarily from close binaries hosting faint or unresolved companions.

\subsubsection{Helium-dominated WDs}
Our analysis is restricted to hydrogen-dominated WDs, and contamination by He-dominated (DB/DZ) WDs could bias the inferred pollution fractions. We have made an effort to remove known He-dominated objects using literature classifications and visual inspection (Section \ref{sec: hecleaning}). However, below $\sim 13,000$ K helium lines become weak or invisible in optical spectra \citep{bergeronComprehensiveSpectroscopicAnalysis2011,koesterDBWhiteDwarfs2015}, making it difficult to reliably distinguish between H- and He-dominated atmospheres. As a result, some WDs classified as DA may in fact have He-dominated atmospheres, which would affect both detectability and abundance determinations.

\subsubsection{Effective temperature distributions}
As discussed in Section \ref{sec: results}, Control Sample 2 and Wide binary sample 2 exhibit different effective temperature distributions. As Ca lines are intrinsically easier to detect in cooler WDs, such differences can lead to different raw detection fractions. For this reason, we rely primarily on the bias-corrected cumulative fraction, which accounts for individual upper limits. As such, hotter stars should not artificially dilute the fraction because they only contribute at abundance levels where Ca would be detectable. Nevertheless, if the intrinsic pollution frequency itself depends on temperature (or WD age), this effect cannot be corrected by our method and could influence the comparison. In addition, temperature differences restrict the abundance range over which the samples can be meaningfully compared.

\subsubsection{Coarse model grid}
The atmospheric model grid used to convert EW into abundances is relatively coarse. This limits the precision of individual [Ca/H] measurements, particularly near the edges of the grid. At low effective temperatures, line broadening and profile changes introduce additional interpolation uncertainties, and in some cases prevent reliable abundance determination, even when a Ca feature is visually present. Moreover, the relationship between EW and abundance is not strictly linear, which further limits abundance precision.

\section{Conclusions} \label{sec: conclusions}

We have presented a comparative study of atmospheric Ca pollution in H-dominated WDs in isolated systems and in wide binary systems, using both low-resolution SDSS spectroscopy and heterogeneous higher-resolution archival data and measurements from different instruments. Our goal was to test whether the presence of a wide stellar companion affects the occurrence rate of outer planetary systems, using WD pollution as an indirect tracer.
Using the homogeneous SDSS-based samples (Control Sample 1 and Wide Binary Sample 1), we find very similar raw pollution fractions for single WDs and WDs in wide binaries. A simple comparison of these fractions would suggest no measurable impact of a wide companion on the incidence of Ca pollution. However, we emphasize that raw fractions alone can be misleading, as the detectability of Ca depends strongly on stellar parameters and data quality.
To address this, we compare bias-corrected cumulative detectability fractions, which account for the Ca abundance that could have been detected in each individual star. With this approach, the cumulative distributions of Ca detections for single and wide-binary WDs are statistically consistent within the uncertainties. This result holds both for the SDSS samples and for the higher-resolution samples, despite the latter exhibiting different raw pollution fractions. The large uncertainties, driven primarily by the small number of detections in the binary samples, limit the strength of quantitative constraints but do not reveal any significant discrepancy between the two populations.
The cumulative detectability approach further shows that the inferred pollution fraction depends on the minimum Ca abundance accessible to a given dataset. As a consequence, studies with different spectral resolutions and signal-to-noise ratios, which are sensitive to different abundance thresholds, can naturally report different overall pollution fractions. This provides a unifying explanation for part of the dispersion in pollution rates reported in the literature. 

Taken at face value, our results indicate that the presence of a wide stellar companion does not measurably alter the probability that a WD hosts remnant outer planetary material. We conclude that the occurrence rate of outer planetary systems in wide binary systems is statistically indistinguishable from that around single stars, based on pollution statistics corrected for detectability biases. This does not imply that wide binaries have no dynamical influence on planetary systems. Rather, multiple processes, such as enhanced early-time perturbations that may deplete reservoirs of small bodies, and late-time perturbations that can promote delivery of material onto the WD, may act in opposite directions, producing similar net pollution rates for single and wide-binary systems.

Finally, the detectability-corrected cumulative fraction framework used here can be readily applied to larger and more homogeneous spectroscopic samples, particularly at higher spectral resolution, to test whether the absence of a detectable difference persists with improved statistics. Such samples will be required to reduce uncertainties and to search for subtle differences as a function of companion separation or Ca abundance level. Studying bias-corrected fractions as a function of detectable abundances may also help constrain the distribution of accreted mass and, indirectly, the size distribution of polluting bodies. Such efforts will be essential to place stronger constraints on the survival and architecture of outer planetary systems through post-main-sequence evolution.

\begin{acknowledgements}
We thank P. Jofré and G. Boggiano for their contributions to this study. C.L.V. and C.A.G. acknowledge support from Agencia Nacional de Investigación y Desarrollo (ANID) through FONDECYT Iniciacion 11230741. C.A.G. acknowledges support from FONDECYT Regular 1262342. A.B. acknowledges the support of a Royal Society University Research Fellowship, URF/R1/211421. C.B.P. acknowledges support from FONDECYT Regular No. 1231057. L.K.R. is Supported by the international Gemini Observatory, a program of NSF NOIRLab, which is managed by the Association of Universities for Research in Astronomy (AURA) under a cooperative agreement with the U.S. National Science Foundation, on behalf of the Gemini partnership of Argentina, Brazil, Canada, Chile, the Republic of Korea, and the United States of America. Based on observations collected at the European Organisation for Astronomical Research in the Southern Hemisphere under ESO programmes 0112.D-0073(A), 167.D-0407(A), 0103.C-0431(B), 71.D-0380(A), 165.H-0588(A), 108.22CP.001, 1103.D-0763(B). 
\end{acknowledgements}

\bibliographystyle{aa}
\bibliography{paper_WDs_sdss.bib}

\onecolumn
\begin{appendix}

\section{Observational details for Wide Binary Sample 2}
\begin{table}[h]
\caption{WDs in Wide Binary Sample 2 observed with higher-resolution instruments. }

\label{table:high_res_observations}      
\centering          
\begin{tabular}{c c c c c c c c} 
\hline\hline       
Gaia EDR3 id & Classification & $T_{\text{eff}}$ & $\log g$ & Instrument & Resolution & SNR & Date obs.\\
- & - & (K) & $(\text{cm s}^{-2})$ & - & - & - & YY-MM-DD\\ 

\hline
1449829295245251968 & DA & 19376 & 8.13 &  & 5400  & 81 & 24-02-17\\ 
2470416951882130560 & DA$^{\star}$ & 11356 & 8.09 &   & 5400  & 97 & 23-10-11 \\ 
2559885724862898304 & DA & 9633  & 8.13 &   & 5400  & 91 &  23-10-15\\ 
2842112183412398336 & DA$^{\dagger}$  & 10603 & 8.28 &  & 5400 & 128 & 23-10-20\\
2844933221011789952 & DA$^{\star}$ & 11871 & 8.03 &   & 5400  & 128 & 23-10-15\\ 
3459790225027896448 & DA$^{\star}$ & 11866 & 8.07 &   & 5400  & 140 & 24-01-21 \\ 
3944621251683633024 & DA$^{\star}$ & 10796 & 8.06 &   & 5400  & 90 & 24-02-14\\ 
4493441175020980480 &  DA$^{\star}$ & 8019 & 7.98 &  & 5400 & 87 & 24-04-06\\
5455210281032856448 &  DA$^{\star}$ & 7329 & 7.53 & X-Shooter & 5400 & 53 & 24-03-13\\
5892401599200616064 & DA$^{\star}$ & 12525 & 8.35 &  & 5400 & 145 & 24-03-16\\
5980497975556687232 & DA$^{\star}$ & 13000 & 8.01 &  & 5400  & 103 & 23-10-04\\ 
6379497207456395776 & DA$^{\star}$ & 10655 & 7.77 &   & 5400  & 81 & 23-10-22 \\ 
6409220477088273920 & DA$^{\star}$ & 13305 & 7.8  &   & 5400  & 192 & 23-12-05 \\ 
6599093628260468224 & DA$^{\star}$ & 9858  & 7.97 &   & 5400  & 108 & 23-10-20\\ 
6655404287348972928 & DA$^{\star}$ & 15814 & 7.89 &   & 5400  & 109 & 23-10-04\\ 
6696618694738631040 & DA$^{\star}$ & 13412 & 7.93 &   & 5400  & 104 & 24-05-11\\ 
6722002157630366720 & DA$^{\star}$ & 11362 & 7.97 &   & 5400  & 95 & 23-10-14\\ 
\hline 
2494399362068211328 & DA$^{\star}$ & 9415 & 8.02 &  & 42000 & 12 & 22-08-10\\
3459790225027896448 & DA$^{\star}$ & 11866 & 8.07 &      & 42000 & 19 & 23-05-31\\ 
3944621251683633024 & DA$^{\star}$ & 10796 & 8.06 &      & 42000 & 14 & 23-06-02 \\ 
4493441175020980480 & DA$^{\star}$ & 8019  & 7.98 & MIKE & 42000 & 4 & 23-07-09 \\ 
5892401599200616064 & DA$^{\star}$ & 12525 & 8.35 &  & 42000 & 10 & 23-06-02 \\ 
5980497975556687232 & DA$^{\star}$ & 13000 & 8.01 &      & 42000 & 19 & 23-06-01 \\ 
6486586685065384576 & DA$^{\star}$ & 9211  & 7.7  &      & 42000 & 4 & 23-07-08\\ 
6813759564048571648 & DA$^{\star}$ & 8745  & 7.88 &      & 42000 & 6 & 23-07-08 \\ 
\hline
\hline
2494399362068211328 & DA$^{\star}$ & 9415 & 8.02 &  X-Shooter & 5400 & 33 & 03-01-16; 03-02-04 \\ 
5909739660590724224 & DA$^{\star}$ & 7403 & 7.98 &  & 5400 & 39 & 19-08-23\\ 
\hline
99915890086770176 & DA$^{\star}$ & 15325 & 8.0 &  & 19540 & 11 & 02-12-31\\ 
2442099751463686528 & DA$^{\star}$ & 17422 & 7.86 &  & 19540 & 22 & 02-07-11; 02-09-13 \\ 
2497895053130247040 & DA$^{\star}$ & 7994 & 7.99 &  & 19540 & 12 & 03-01-16; 03-02-04\\ 
2610488514148351360 & DA$^{\star}$ & 6611 & 7.94 &  & 19540 & 11 & 00-07-12; 00-07-17; 03-07-03\\ 
3083590225639363712 & DA$^{\star}$ & 9200 & 7.9 &  & 19540 & 17 & 22-01-16 \\ 
3201594074140023936 & DA$^{\star}$ & 8105 & 7.95 &  & 19540 & 10 & 21-11-01\\ 
3459790225027896448 & DA$^{\star}$ & 11866 & 8.07 & UVES & 19540 & 17 & 22-01-16\\ 
3861237313489649280 & DA$^{\star}$ & 13410 & 7.38 &  & 19540 & 19 & 22-01-16\\ 
4460086458999242368 & DA$^{\star}$ & 16976 & 7.86 &  & 19540 & 26 & 02-08-04; 03-07-19\\ 
4488750108662714624 & DA$^{\dagger}$ & 10575 & 8.25 &  & 19540 & 16 & 02-07-17; 07-27; 03-03-22\\ 
4642153627368210560 & DA$^{\star}$ & 13009 & 7.91 &  & 19540 & 16 & 21-10-10\\ 
5499525444356625408 & DA$^{\star}$ & 16257 & 7.89 &  & 19540 & 20 & 21-10-21\\ 
5544588928434374016 & DA$^{\star}$ & 16500 & 7.94 &  & 19540 & 28 & 21-10-23\\ 
5892401599200616064 & DA$^{\star}$ & 12525 & 8.01 &  & 19540 & 11 & 22-01-17\\ 
6009537829925128064 & DA$^{\star}$ & 10391 & 8.02 &  & 19540 & 38 & 00-07-14; 03-05-12; 22-01-24\\ 

\hline
\hline                  
\end{tabular}
\tablefoot{Values of $T_{\text{eff}}$ and $\log g$ were adopted from \cite{gentilefusilloCatalogueWhiteDwarfs2021}. The SNR was estimated in the vicinity of the Ca II K line. Spectral classifications marked with $^{\star}$ and $^{\dagger}$ were taken from \cite{vincentClassificationParameterizationLarge2024} and \cite{caronSpectrophotometricAnalysisCool2023}, respectively; classifications for the remaining objects were assigned in this work. Targets listed below the double horizontal line are taken from the archive. Objects observed multiple times have not been merged or removed.}
\end{table}

\newpage
\section{Upper Limits} \label{app:UL}

\begin{table}[h]
 \caption[]{\label{table:A_upperlimits_singles}  Example of the table containing upper limits for single WDs in the SDSS sample (Control Sample 1).}

 \centering
\begin{tabular}{ccccccc}
 \hline \hline
  Gaia EDR3 id & $T_{\text{eff}}$ (K) & $\log g\,\,\text{(cm s}^{-2})$ & SNR & EW (m\AA) & $\log{\text{(Ca/H)}}$\\ %& 
  \hline
  152935195517952 & 12593 & 7.65 & 33 & $<489$ & $<-7.74$ \\
  288175125714560 & 12615 & 7.91 & 26 & $<551$ & $<-5.60$ \\
  298895364329216 & 16672 & 8.38 & 42 & $<360$ & $<-5.00$ \\
  3143095722559872 & 18068 & 7.89 & 39 & $<433$ & $<-4.50$ \\
  3250362530184064 & 10855 & 7.87 & 18 & $<789$ & $<-6.02$ \\
  3357427475070208 & 10535 & 7.84 & 11 & $<1289$ & $<-5.71$ \\
  3752804984535936 & 16114 & 7.79 & 15 & $<988$ & $<-4.06$ \\
  4404647876519040 & 9230 & 7.87 & 10 & $<1301$ & $<-6.76$ \\
  5253466557760640 & 12134 & 8.07 & 24 & $<644$ & $<-5.74$ \\
  6918986155511040 & 12004 & 8.01 & 23 & $<712$ & $<-5.68$ \\
  ... & ... & ... & ... & ... & ...\\
  ... & ... & ... & ... & ... & ...\\
  
 \hline
\end{tabular}
\tablefoot{The values of $T_{\mathrm{eff}}$ and $\log g$ were obtained from \cite{gentilefusilloCatalogueWhiteDwarfs2021}. The reported SNR was estimated in the vicinity of the Ca~II~K line, and the corresponding $\log(\mathrm{Ca/H})$ upper-limit abundances are listed.}
\end{table}

\begin{table}[h]
 \caption[]{\label{table:A_upperlimits_widebinaries}  
 Example of the table containing upper limits for Wide Binary Sample 1, from SDSS.}
 \centering
\begin{tabular}{ccccccc}
 \hline \hline
  Gaia EDR3 id & $T_{\text{eff}}$ (K) & $\log g\,\,\text{(cm s}^{-2})$ & SNR & EW (m\AA) & $\log{\text{(Ca/H)}}$\\ 
  \hline
  7033270940879744 & 12427 & 8.05 & 23 & $<689$ & $<-5.54$ \\
  78023441051581568 & 18021 & 7.99 & 12 & $<1241$ & $<-4.00$ \\
  95188672986669824 & 8307 & 7.37 & 12 & $<1105$ & $<-7.24$ \\
  126595810317176448 & 18110 & 8.27 & 11 & $<1213$ & $<-4.00$ \\
  143563478052546688 & 10037 & 8.07 & 24 & $<672$ & $<-6.72$ \\
  235840781689958016 & 19737 & 8.30 & 10 & $<1265$ &  \\
  290410535910963584 & 16405 & 7.82 & 27 & $<558$ & $<-4.54$ \\
  294902659385996288 & 7542 & 7.74 & 14 & $<1054$ & $<-8.01$ \\
  299129182083300352 & 10312 & 7.63 & 12 & $<1264$ & $<-5.80$ \\
  304311798860298880 & 7135 & 7.92 & 10 & $<1402$ & $<-8.13$ \\
  ... & ... & ... & ... & ... & ...\\
  ... & ... & ... & ... & ... & ...\\
  
 \hline
\end{tabular}
\tablefoot{The values of $T_{\mathrm{eff}}$ and $\log g$ were obtained from \cite{gentilefusilloCatalogueWhiteDwarfs2021}. The reported SNR was estimated in the vicinity of the Ca~II~K line, and the corresponding $\log(\mathrm{Ca/H})$ upper-limit abundances are listed.}
\end{table}

\begin{table}[h]
 \caption[]{\label{table:B_upperlimits}  
 Example of the table containing upper limits for Wide Binary Sample 2.}

 \centering
\begin{tabular}{cccccccc}
 \hline \hline
  Gaia EDR3 id & $T_{\text{eff}}$ (K) & $\log g\,\,\text{(cm s}^{-2})$ & SNR & EW (m\AA) & $\log{\text{(Ca/H)}}$ & Instrument\\  
  \hline
  99915890086770176 & 15325 & 8.00 & 11 & $<60$ & $<-7.38$ & UVES\\
  2442099751463686528 & 17422 & 7.86 & 22 & $<33$ & $<-7.58$ & UVES\\
  2497895053130247040 & 7994 & 7.99 & 12 & $<56$ &  & UVES\\
  2610488514148351360 & 6611 & 7.94 & 11 & $<62$ &  & UVES\\
  3083590225639363712 & 9200 & 7.90 & 17 & $<42$ & $<-10.00$ & UVES\\
  3201594074140023936 & 8105 & 7.95 & 10 & $<73$ & $<-10.00$ & UVES\\
  3459790225027896448 & 11866 & 8.07 & 17 & $<42$ & $<-8.75$ & UVES\\
  3861237313489649280 & 13410 & 7.38 & 19 & $<38$ & $<-8.5$ & UVES\\
  4460086458999242368 & 16976 & 7.86 & 26 & $<28$ & $<-7.98$ & UVES\\
  4488750108662714624 & 10575 & 8.25 & 16 & $<45$ & $<-9.34$ & UVES\\
  ... & ... & ... & ... & ... & ... & ...\\
  ... & ... & ... & ... & ... & ... & ...\\
  
 \hline
\end{tabular}
\tablefoot{The values of $T_{\mathrm{eff}}$ and $\log g$ were obtained from \cite{gentilefusilloCatalogueWhiteDwarfs2021}. The reported SNR was estimated in the vicinity of the Ca~II~K line, and the corresponding $\log(\mathrm{Ca/H})$ upper-limit abundances are listed.}
\end{table}

\end{appendix}

\end{document}